# Thermal denaturation of a protein (CoVE) by a coarse-grained Monte Carlo simulation

R.B. Pandey
School of Mathematics and Natural Sciences, University of Southern Mississippi, Hattiesburg, MS 39406-5043, USA
Email: ras.pandey@usm.edu

**Abstract:** Thermal response of a protein (CoVE) conformation is studied by a coarse-grained Monte Carlo simulation. Three distinct segments, the N-terminal, Trans-membrane, and C-terminal are verified from its specific contact profile. The radius of gyration (Rg) is found to exhibit a non-monotonic sub-universal thermal response: Rg decays on heating in native phase (low-temperature regime) in contrast to a continuous increase on further raising the temperature before its saturation to a random-coil in denature phase. The globularity index (a measure of effective dimension) of the protein decreases as the protein denatures from a globular to a random-coil conformation.

**Introduction**

The outbreak of COVID 19 pandemic has caused an urgent need to understand a wide range of issues, both basic and applied, from epidemiology to diverse multi-scale molecular structures of its constitutive elements of the novel corona virus (CoV-2) and its interaction with surroundings [1−4]. The RNA genome of CoV-2 is believed to encode only four major structural proteins (out of 29): spike (S), nucleocapsid (N) protein, membrane (M) protein, and the envelope (E) protein [4, 5]. These structural proteins are key components in assembly of a viral particles. The envelope protein CoVE is the smallest with 76 residues ($^1M^2Y\ldots^{76}V$) [6] but plays an important role in life cycle of the virus such as formation of envelope as an integral membrane protein in viral assembly, budding and pathogenesis [4−12]. We would like to focus on the folding dynamics of CoVE, which has not been analyzed to our knowledge as presented here.

Sequence of CoVE protein and its specific characteristics such as formation of membrane ion channel, virulence intensity, etc. are studied in depth [4−12]. Three distinct segments of the protein starts with the hydrophilic amino (N) terminal with about 10 residues ($^1M - {}^{10}G$), followed by the hydrophobic transmembrane (TM) segment of about 25-27 residues ($^{11}T - {}^{37}L$), and ends with a relatively large hydrophilic carboxyl terminal consisting of the remaining residues ($^{38}R - {}^{76}V$). The structural details of how these segments interact, compete and cooperate is not fully understood. The main objective of this article is to explore the effect of temperature on the structural dynamics by analyzing some of the local and global physical quantities by a coarse-grained Monte Carlo simulation. The model and method in described in brief next followed by results and discussion and a conclusion.

**Model and method**

The model adopted here has been already described and used to investigate structural dynamics of a number of proteins [13−16]. In coarse-grained description, CoVE is represented by a chain of 76 residues [6] in a specific sequence on a cubic lattice with ample degrees of freedom for each residue to move and their covalent bonds to fluctuate. A residue is represented by a cubic node of size $(2a)^3$ where $a$ is the lattice constant. Consecutive nodes are tethered together by flexible covalent bonds; the bond length between consecutive nodes varies between *2* and $\sqrt{10}$ in unit of lattice constant ($a$). Each residue interacts with surrounding residues within a range ($r_c$) with a generalized Lennard-Jones potential,

$$U_{ij} = \left[ |\varepsilon_{ij}| \left(\frac{\sigma}{r_{ij}}\right)^{12} + \varepsilon_{ij} \left(\frac{\sigma}{r_{ij}}\right)^{6} \right], \; r_{ij} < r_c \qquad (1)$$

where $r_{ij}$ is the distance between the residues at site $i$ and $j$; $r_c = \sqrt{8}$ and $\sigma = 1$ in units of lattice constant. The potential strength $\varepsilon_{ij}$ in phenomenological interaction (1) is based on a knowledge-based [17] residue-residue contact matrix which has been developed over many years [18 −22] from a huge and growing ensemble of protein structures in PDB. Each residue performs its stochastic movement with the Metropolis algorithm, i.e. with the Boltzmann probability *exp(-ΔE/T)* where *ΔE* is the change in energy between new and old position subject to excluded volume constraints. As usual, attempts to move each residue once defines unit Monte Carlo time step. Physical quantities are measured in arbitrary unit i.e. length in unit of lattice constant and the temperature *T* in reduced units of the Boltzmann constant.

All simulations presented here are performed on a *$150^3$* cubic lattice for a range of temperature *T = 0.010-0.030* with *100* independent samples; different sample sizes are also used to make sure that the qualitative results are independent of finite size effects. Both local and global physical quantities such as radius of gyration, root mean square displacement of the center of mass, structure factor, contact map, etc. are examined as a function of temperature.

**Results and discussion**

Snapshots of the protein (CoVE) at representative temperatures ($T=0.020, 0.024, 0.030$) presented in figure 1 illustrate the conformation changes as it denatures on raising the temperature. Obviously the protein conforms to a globular (compact) structure at low temperature ($T=0.020$) and expands as it denatures ($T=0.030$). The segmental globularity in the transmembrane segment of the protein is retained at least in part even at the high temperature (see below). Note that snapshots do not represent the average conformation; it is an illustrative mean that needs to be quantified. Therefore, some quantitative measures of segmental globularity are highly desirable. Average number ($N_n$) of residues within the range of interaction of each along the contour of the protein may provide some insight into the local restructuring as a function of temperature.

Figure 2 shows the contact profiles at representative temperatures. At low temperatures ($T=0.012 – 0.020$, in native phase), the coagulating centers are distributed sporadically particularly in N- and C-terminal regions along the backbone (e.g. $^1M^2Y$, $^{12}L^{13}I^{14}V$ (in N-terminal), $^{51}L^{52}V^{53}K$, $^{56}V^{57}Y^{58}V^{59}Y$, $^{74}L^{75}L^{76}V$ (in C-terminus)) with relatively large $N_n$. The transmembrane segment ($^{18}L – ^{44}C$), however, has the largest globular cluster with high number $N_n$ of surrounding residues. Clearly, there is no significant change in contact profiles in native phase. However, raising the temperature ($T=0.024$) leads to a substantial decrease in in segmental globularity in both N- and C-terminals with low contact number $N_n$ while retaining some globularity in TM segment with a relatively large $N_n$ (see figure 2). Thus, it is easy to identify three distinct (N- and C-terminals separated by TM) segments from the evolution of contact profiles with the temperature.

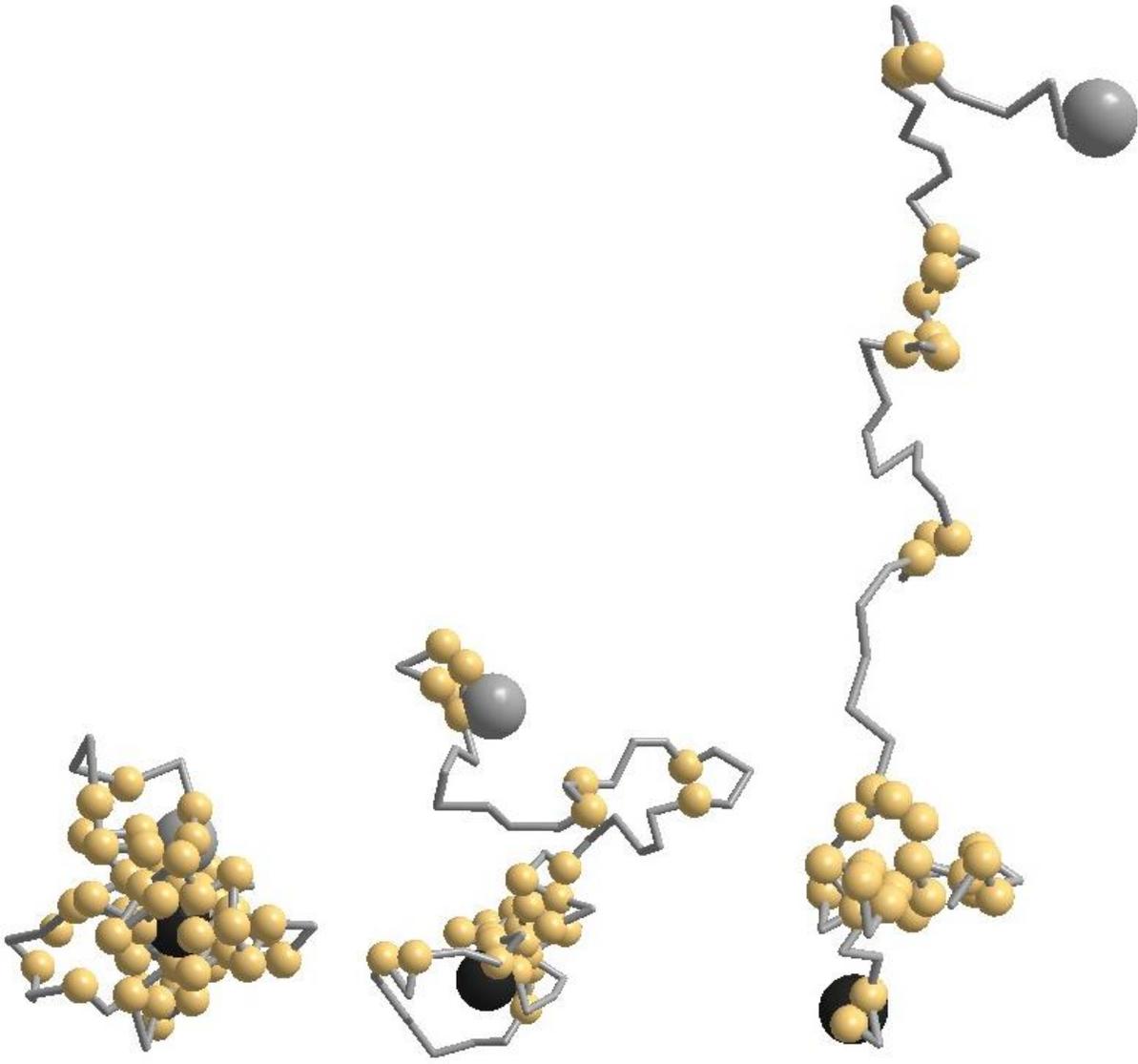

Figure 1: Snapshot of CoVE protein at the end of $10^7$ MCS time. Large black sphere represents the first residue ($^1$M) and grey, the last ($^{76}$V) of the protein. Smaller golden spheres represent the presence of other residues within the range of interaction excluding the consecutively connected residues. Backbone grey lines are covalent bonds connecting the residues. Snapshot at the left (*T=0.020*), center (*T=0.024*), right (*T=0.030*) are at the representative temperatures to illustrate denaturing of the protein.

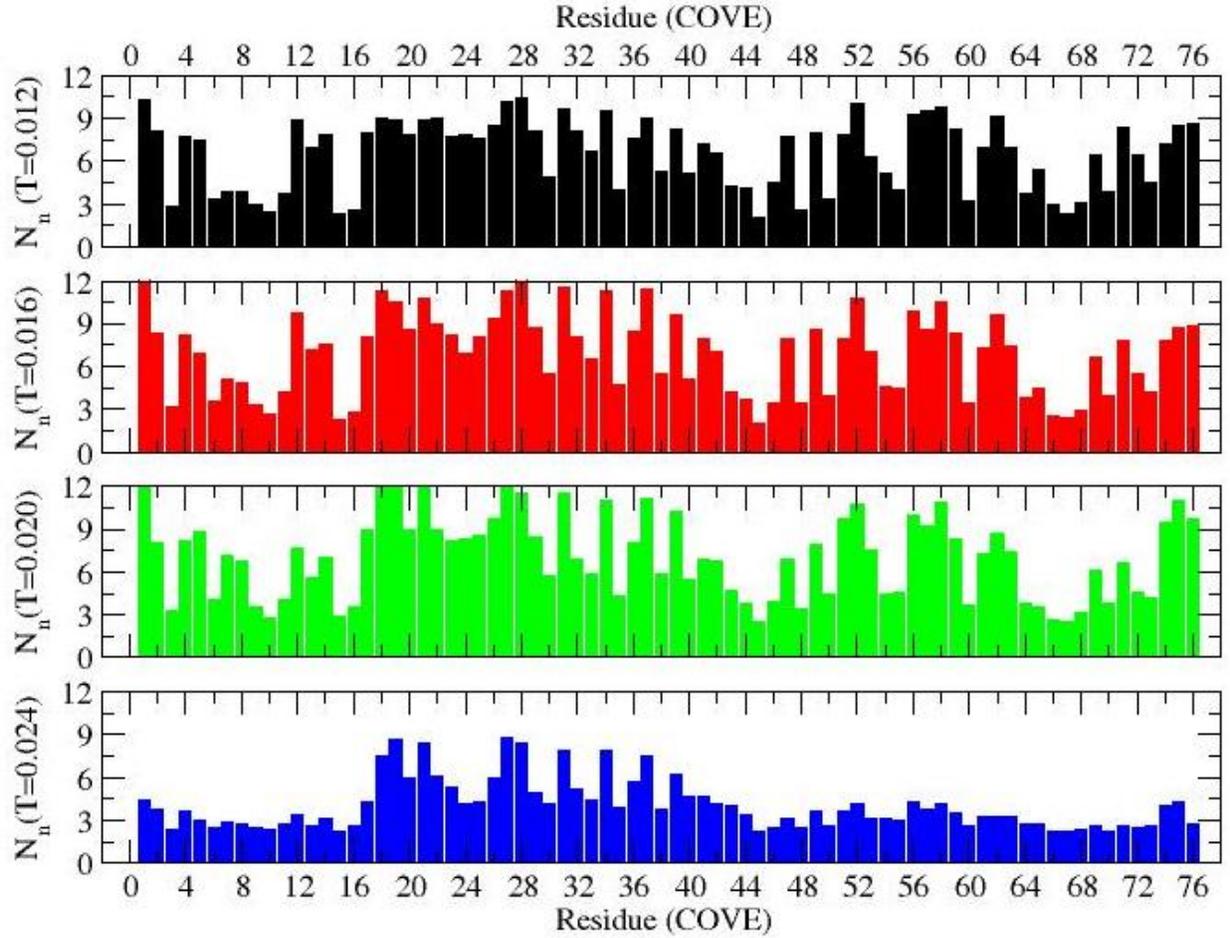

Figure 2: Average number ($N_n$) of residues around each within the range of interaction at a range of temperature ($T=0.012 - 0.024$).

The global structure and dynamics of the protein emerge from the stochastic movement of each residues and their configurational stabilities. Variations of the root mean square displacement of the center of mass ($R_c$) with the time step ($t$) at a range of temperature presented in figure 3 show, how the asymptotic dynamics depends on temperature. One can identify the type of dynamics by estimating the power-law exponent ($v$) by fitting the data with a scaling, $R_c \propto t^v$ where $v = ½$ represents diffusion. Accordingly, the protein is found to move extremely slow at low temperatures with ultra-sub-diffusive dynamics (*i.e. v = 0.05 T=0.020*) and resumes diffusion at higher temperatures (*i.e. v = 0.47,* at *T=0.024*, see figure 3).

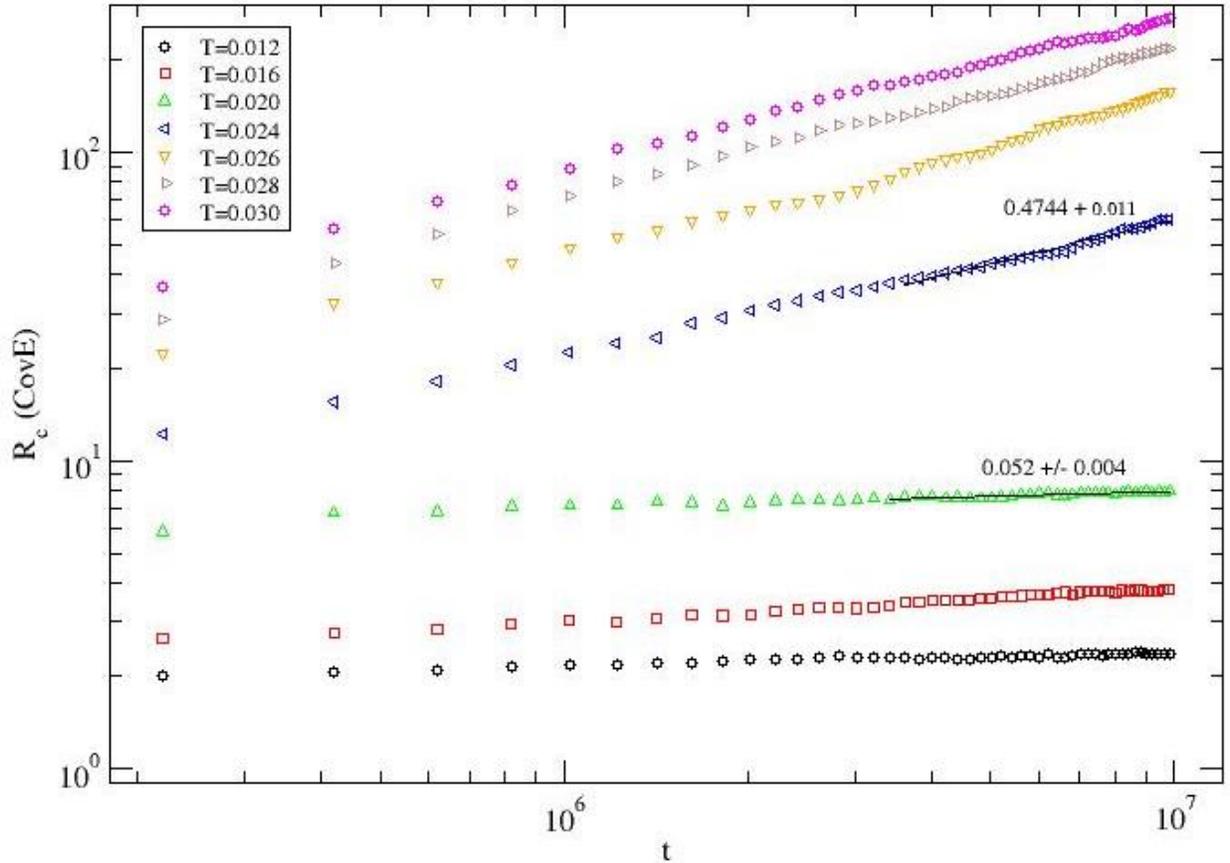

Figure 3: Variation of the average root mean square displacement (RMSD) of the center of mass ($R_c$) of the protein with the time step ($t$) for a range of temperature ($T=0.012 - 0.030$) on a log-log scale. Slopes of the fitted data (solid black lines) in the asymptotic regime at a low ($T=0.012$) and a high ($T=0.024$) temperature are included with corresponding estimates.

The radius of gyration ($R_g$) of the protein is a measure of its average size. Variation of the average radius of gyration ($R_g$) with the temperature ($T$) is presented in figure 4. In contrast to denature phase ($T=0.020 - 0.030$) where the radius of gyration ($R_g$) increases on raising the temperature, it decays on increasing the temperature in the native phase ($T=0.010 - 0.020$). Such an opposite thermal response in native and denature phases is recently observed in other membrane proteins in recent years [23, 24]. In order to claim such thermal response, a universal or sub-universal characteristics for the membrane proteins, more studies on many different membrane proteins are needed. The thermal response of the nucleocapsid protein (CoVN) in non-monotonic but very different from that of CoVE. We hope this results will stimulate further investigations in this direction.

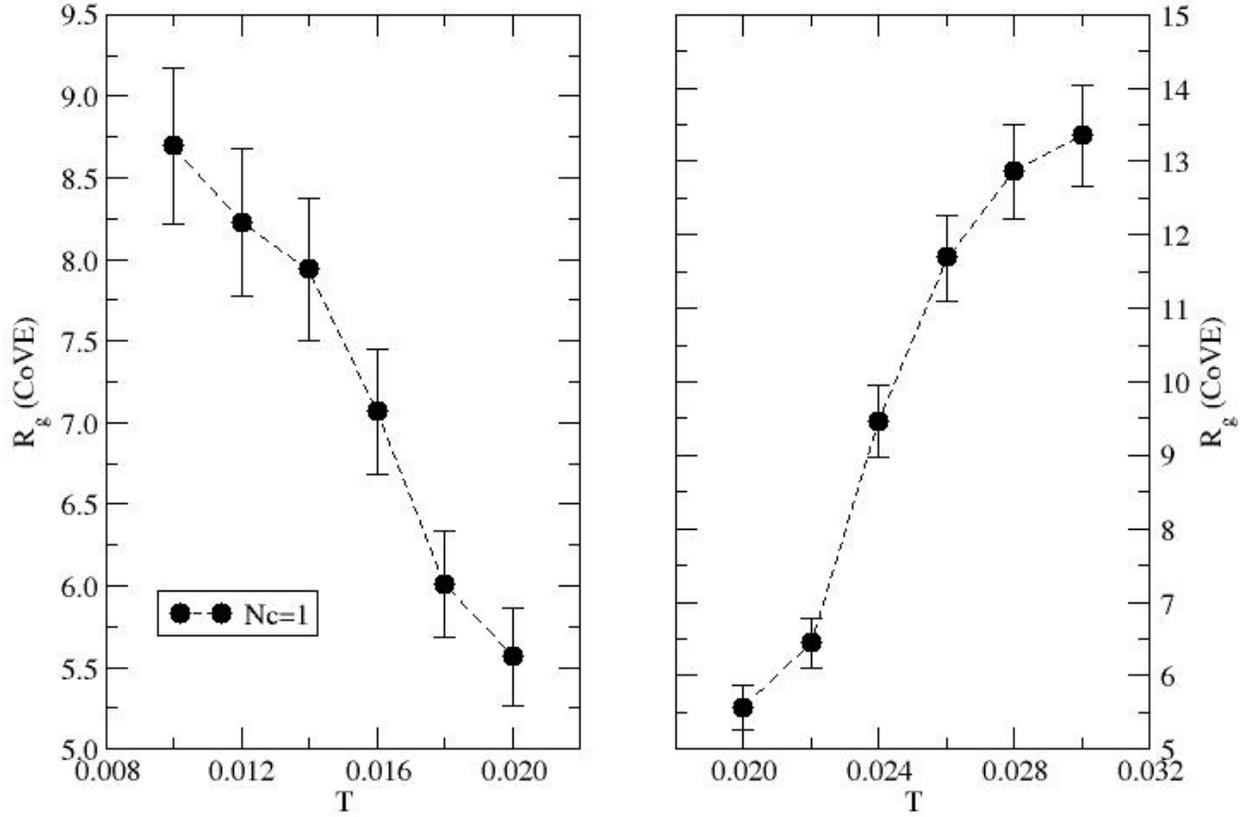

Figure 4: Average radius of gyration ($R_g$) of CoVE versus temperature.

Analysis of the structure factor can also provide insight into the global conformation of the protein. The structure factor ($S(q)$) is defined as,

$$S(q) = \langle \frac{1}{N} \left| \sum_{j=1}^{N} e^{-i\vec{q} \cdot r_j} \right|^2 \rangle_{|\vec{q}|} \qquad (2)$$

where $r_j$ is the position of each residue and the magnitude of the wave vector ($q$) of wavelength $\lambda$ is given by $|q| = 2\pi/\lambda$. Using a power-law scaling for the structure factor with the wave vector, $S(q) \propto q^{-1/\gamma}$, one can evaluate the exponent $\gamma$. Variations of the structure factor with the wavelength (comparable to radius of gyration) is presented in figure 5 for representative temperatures.

In general, the radius of gyration ($R_g$) of a polymer chain shows a power-law scaling with number ($N$) of monomer (residues), $R_g \propto N^x$ with a well-defined exponent $x$. Since a protein is a hetero-polymer chain of residues, one may use the same scaling law here for the protein with the radius of gyration comparable to wavelength ($R_g \sim \lambda$). Then the power-law exponent $x = \gamma$ and one may be able to estimate the globularity index $g$ which is the effective dimension ($D$) for the spread of the protein ($g \cong D$). Since, $N \propto R_g^D$, $D = 1/\gamma$. Higher magnitude of $D$ represents more compact structure with higher degree of globularity. Note that this scaling analysis is an estimate and should not be taken as a precise measurement. Obviously $D$ cannot be higher than 3, however high value indicates more compact conformations. Clearly, the protein remains globular (three dimensional) at low temperatures in native phase and it denatures into a ramified random coil (two dimesnional) structures at high temperatures (see figure 5).

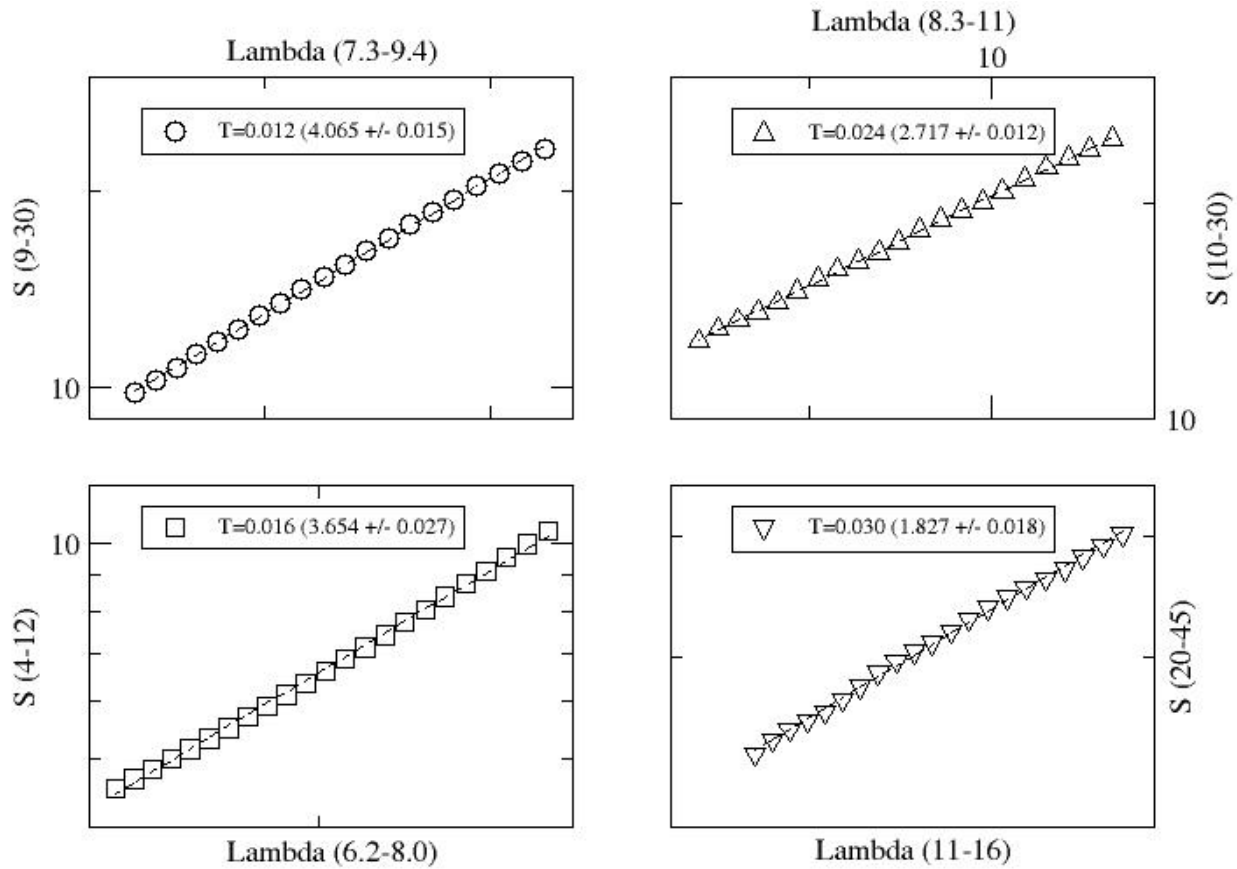

Figure 5: Structure factor (S) of the protein versus wave length (lambda) at representative low and high temperatures. Data points are selected with wavelength comparable to radius of gyration of the protein at corresponding temperatures with the estimates of slopes. The number along the axes are the range of the axes (starting and end points).

**Conclusion**

Monte Carlo simulations are performed with a coarse-grained model of envelope membrane protein CoVE of the novel corona virus to assess its structural variability as a function of temperature. The evolution of the protein's conformation as a function of temperature is clearly seen in visualization via snapshots and animations. Segmental organizing of residues is quantified by examining local physical quantities such as contact and mobility profiles as a function of temperature. From the variations in contact profiles as a function of temperature, it is feasible to identify three distinct segments (N-terminal and C-terminal separated by a transmembrane segment) which is consistent with the previous studies on CoVE [4-12]. The radius of gyration exhibits a non-monotonic thermal response: the protein CoVE expands on heating in native phase and expands continuously on raising the temperature before reaching a saturation in denatured phase. Such a unique non-monotonic thermal response has been recently observed in other membrane proteins (unrelated to coronavirus). This leads to speculate that thermal response may be a unique characteristic to identify different classes of proteins – obviously more studies are needed to confirm this observation. Thermal response of CoVE is different from another structural protein CoVN (a nucleocapsid protein) of the novel coronavirus [25]. Variation of the structure factor with the wave vector is studied in detail for a wide range of temperature. Scaling analysis of the structure factor provides a quantitative estimate of the overall globularity with an effective dimension ($D$). CoVE is found to remain globular ($D \sim 3$) in its native phase and unfolds continuously to a random coil ($D \sim 2$) in denature phase on raising the temperature.